\documentclass[aps,prr,twocolumn,floatfix,showpacs,citeautoscript,longbibliography,hyperlinks]{revtex4-2}
\usepackage{amsmath}
\usepackage{graphicx,graphics}
\usepackage{epstopdf}
\usepackage{amsmath}
\usepackage{amssymb}
\usepackage{amsfonts}
\usepackage[colorlinks=true,citecolor=blue]{hyperref}
\usepackage{etoolbox}
\apptocmd{\sloppy}{\hbadness 10000\relax}{}{}

\usepackage{xcolor}
\newcommand{\revision}[1]{{{#1}}}

\newcommand{\revisiontwo}[1]{{{#1}}}

\begin{document}
\title{Effective temperature pulses in open quantum systems}
\author{Pedro Portugal}
\author{Fredrik Brange}
\author{Christian Flindt}
\affiliation{Department of Applied Physics, Aalto University, 00076 Aalto, Finland}

\begin{abstract}
Controlling the temperature of nano-scale quantum systems is becoming increasingly important in the efforts to develop thermal devices such as quantum heat valves, heat engines, and refrigerators, and to explore fundamental concepts in quantum thermodynamics. In practice, however, it is challenging to generate arbitrary time-dependent temperatures, similarly to what has been achieved for electronic voltage pulses. To overcome this problem, we here propose a fully quantum mechanical scheme to control the time-dependent environment temperature of an open quantum system. To this end, we consider a collection of quantum harmonic oscillators that mediate the interactions between the quantum system and a thermal reservoir, and we show how an effective time-dependent temperature can be realized by modulating the  oscillator frequencies in time. By doing so, we can apply effective temperature pulses to the quantum system, and it can be cooled below the temperature of the environment. Surprisingly, the scheme can be realized using only a few oscillators, and our proposal thereby paves the way for controlling the temperature of open quantum systems.
\end{abstract}

\maketitle

\section{Introduction} 
Temperature is a central concept in thermodynamics, which characterizes the tendency of a physical system to emit or absorb heat~\cite{callen1998thermodynamics}.  Typically, temperatures vary only slowly in time, however, fast temperature modulations are playing an increasingly important role in quantum thermodynamics, for instance, in relation to finite-time thermal machines~\cite{blickle2012realization,asingleatom,martinez2016brownian,PhysRevA.99.022105,PhysRevLett.114.120601,PhysRevResearch.2.043262}, fluctuation theorems~\cite{PhysRevLett.92.230602,PhysRevLett.106.200602,cuetara2014,PhysRevLett.116.068301,rademacher2021nonequilibrium}, and quantum calorimetry~\cite{QuantCalPhysicsToday,PhysRevB.98.205414,Karimi2020,Pekola:2021}. Recently, the field of thermotronics has emerged with the goal of realizing thermal capacitors~\cite{Wang:2007,Dhar:2008,Li:2012,portugal2021heat}, transistors \cite{Segal:2005,Ojanen:2008,Ruokola:2009}, and other nonlinear  components such as thermal memristors~\cite{diventra2009,Wu:2009} that are driven by time-dependent temperatures instead of voltages. However, fast control of the temperature at the nanoscale is challenging, and most proposals for temperature control~\cite{Solano2010,PhysRevLett.109.103603,PhysRevE.87.032159,PhysRevE.90.032116,Petrosyan2014,Gieseler2015,rademacher2021nonequilibrium} rely on  semiclassical techniques or time-dependent couplings to environments at different temperatures~\cite{brandner2015,PhysRevE.93.062134}.

Several approaches to temperature control are based either on random forces acting on a system~\cite{Petrosyan2014,Solano2010,PhysRevE.90.032116,gieseler2018} or on the use of feedback cooling (or heating) by a parametric drive~\cite{Gieseler2015,PhysRevLett.109.103603,PhysRevE.87.032159,rademacher2021nonequilibrium}. As an example, the stiffness of an optical potential may be changed in response to the movement of a trapped particle. More generally, structured environments have been employed to generate non-Markovian dynamics, for instance, by using a finite-size system that mediates the interactions between an open quantum system and its environment~\cite{garraway2009}. Alternatively, a finite-size system can mimic a macroscopic environment for a limited period of time~\cite{PhysRevB.92.155126,PRXQuantum.2.010340}. Potentially, these methods may also be used to control the effective  temperature of an environment and thereby apply temperature pulses to an open quantum system as we will see.

\begin{figure}[b!]
    \centering
\includegraphics[width=0.95\columnwidth]{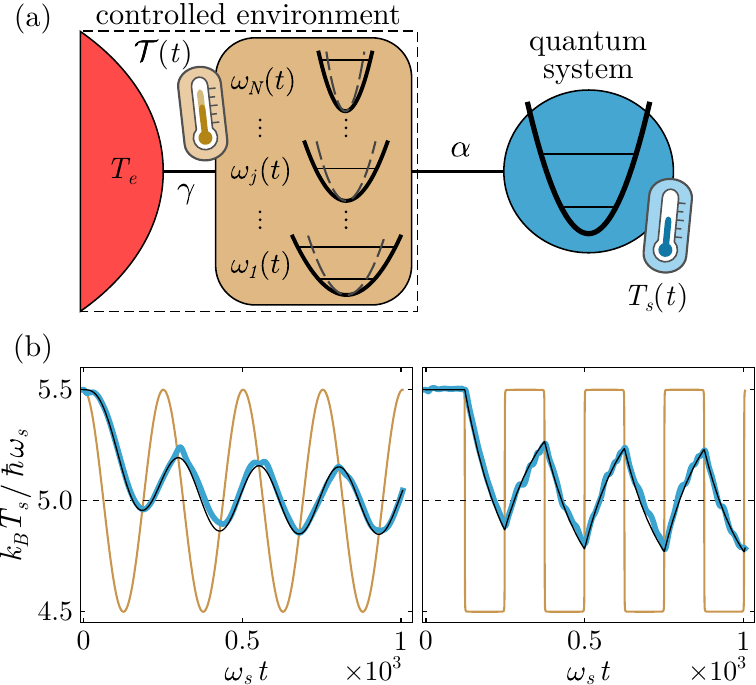}
    \caption{Effective temperature pulses. (a) A quantum system (blue) is coupled with strength~$\alpha$ to an environment consisting of $N$ quantum harmonic oscillators (brown) that are coupled with the rate $\gamma$ to a thermal reservoir (red) at the fixed temperature~$T_e$. We modulate the oscillator frequencies, $\omega_j(t)$, to generate a time-dependent effective temperature $\mathcal{T}(t)$ that the quantum system experiences. (b) As an application, we take a single harmonic oscillator with the frequency $\omega_s$ as the quantum system,  and we show its temperature $T_s(t)$ (blue) in response to the temperature pulses $\mathcal{T}(t)$ (brown) with $N=16$. As a comparison, we show the time-dependent temperature obtained from a Lindblad equation for the quantum system only with $\mathcal{T}(t)$ inserted (black)\revision{, see Eq.~(\ref{eq:effLind}).} The parameters are $\alpha=0.05\nu\omega_s$, $\gamma=0.01\omega_s$, and $k_B T_e=5\hbar \omega_s$.}
    \label{fig:system}
\end{figure}

In this paper, we propose and analyze a fully quantum mechanical scheme to control the effective temperature of an open quantum system. Figure~\ref{fig:system}(a) illustrates our setup which consists of a collection of quantum harmonic oscillators that mediate the interactions between a small quantum system and a thermal reservoir, which is kept at a fixed temperature. Inspired by Luttinger, we modulate the frequencies of the oscillators in time to generate a time-dependent effective temperature that the quantum system is exposed to~\cite{Luttinger:1964,Eich:2014}. In particular, we determine how the oscillator frequencies must be modulated in time to realize a desired effective temperature profile $\mathcal{T}(t)$ as illustrated in panel (b), where we show how the temperature of the quantum system responds to two specific temperature drives using $N=16$ quantum harmonic oscillators. This approach also allows us to cool the quantum system and maintain its temperature below that of the environment. Our work thereby provides a practical strategy for controlling the effective temperatures of open quantum systems, and it opens up avenues for further theoretical and experimental investigations of temperature-driven quantum systems.

\section{Effective temperatures}
The Hamiltonian of the quantum system and the harmonic oscillators reads
\begin{equation}
\hat H(t)=\hat H_s+\hat H_a(t)+ \hat H_i,
\label{eq:Ham}
\end{equation}
where the quantum system itself is a harmonic oscillator $\hat H_s=\hbar\omega_s (\hat a_{s}^\dagger \hat a_{s}+1/2)$ that  interacts with the other oscillators, $\hat H_a(t)=\sum_{j=1}^{N} \hbar \omega_{j}(t) (\hat a_j^\dagger \hat a_j+1/2)$,  through the coupling $\hat H_i=\hbar\alpha\sum_{j=1}^{N}  (\hat a_s^\dagger \hat a_j + \hat a_j^\dagger \hat a_s)$, and $\hat a^\dagger_{s,j}$ and  $\hat a_{s,j}$ are the usual ladder operators. Our scheme applies to other types of quantum systems, such as two-level systems, however, it is useful to consider an oscillator, since it can be described by an effective temperature that we denote by $T_s(t)$. The key idea is now to modulate the frequencies $\omega_j(t)$ in time to generate an effective temperature profile $\mathcal{T}(t)$ that the quantum system experiences. 

\revisiontwo{The coupling between the quantum system and the oscillators is weak, $\alpha\ll\omega_s,\omega_j$. In addition, the oscillators are weakly coupled to a heat bath at the temperature $T_e$, and the driving is slower than the bath correlation time. The reduced density matrix $\hat \rho(t)$ of system and oscillators then evolves according to the Lindblad equation~\cite{breuer2002theory,Rivas:2012,Albash:2012}
\begin{equation}\label{eq:lindblada}
    \frac{\mathrm d \hat \rho (t)}{\mathrm d t}= -\frac{i}{\hbar}[\hat H(t),\hat \rho(t)] + \sum_{j=1}^{N}\gamma_j(\omega_j)  \mathcal D_j\hat \rho(t),
\end{equation}
where we have introduced the dissipators $\mathcal{D}_j\hat\rho= [1+n_{T_e}(\omega_j)](\hat a_j \hat \rho \hat a_j^\dagger-\frac{1}{2}\{ \hat a_j^\dagger \hat a_j,\hat \rho\})+n_{T_e}(\omega_j)(\hat a_j^\dagger \hat \rho \hat a_j-\frac{1}{2}\{\hat a_j \hat a_j^\dagger,\hat \rho\})$, the rates $\gamma_j$ depend on the oscillator frequencies via the spectral function of the heat bath, and $n_{T}(\omega)=\left[\exp(\hbar\omega/k_BT)-1\right]^{-1}$ is the Bose-Einstein distribution. The derivation of Eq.~(\ref{eq:lindblada}) is provided in App.~\ref{appA} together with the relevant assumptions. We note that the oscillator frequencies in the Hamiltonian may be slightly renormalized due to a small Lamb-shift.}
 
\section{Driving protocol} 
To find the driving protocol for the oscillator frequencies $\omega_j(t)$ to realize a given temperature profile $\mathcal{T}(t)$, we first set the coupling to zero ($\alpha=0$) and impose that the $N$ oscillators remain in the thermal state 
\begin{equation}
\label{eq:effectivethermal}
    \hat\rho_a(t)=e^{-\hat H_a(t)/k_B\mathcal T (t)}/Z(t)
\end{equation}
with $Z(t)$ defined so that $\mathrm{tr}\{\hat\rho_a(t)\}=1$. With $\mathcal T (t)=T_e$, this is the state of the oscillators in equilibrium with the reservoir. Generally, however, it is a non-equilibrium state, but it takes the form of an equilibrium density matrix with the effective temperature $\mathcal T (t)$. Inserting Eq.~\eqref{eq:effectivethermal} into Eq.~\eqref{eq:lindblada}, we obtain the differential equation 
\begin{equation}
\frac{\mathrm d}{\mathrm dt} n_{\mathcal{T}}(\omega_j)= \gamma(\omega_j) \left[n_{T_e}(\omega_j)-n_{\mathcal{T}}(\omega_j)\right]
\label{eq:frequencies}
\end{equation}
for each oscillator frequency, see App.~\ref{appB}. The left-hand side is the required change in the population of each oscillator to retain the state in Eq.~\eqref{eq:effectivethermal}, and the right-hand side is the net absorption rate of energy quanta from the environment. Without the  reservoir ($\gamma_j=0$), we easily find $\omega_j(t)/\omega(0) = \mathcal{T}(t)/\mathcal{T}(0)$, showing that the change in the frequencies is directly proportional to the desired temperature profile. Furthermore, for $\mathcal{T}(t) = T_e$, we find $\dot{\omega}_j = 0$, and no driving is needed, as one would expect. 
	
\begin{figure}
  \includegraphics[width=0.95\columnwidth]{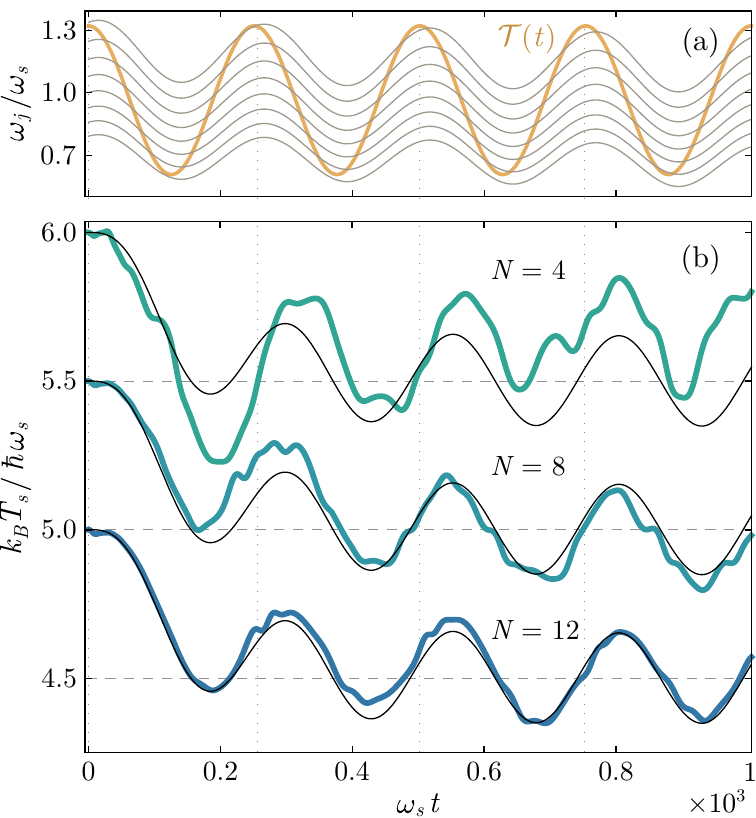}
  	\caption{System temperature for different numbers of oscillators. (a) We show the oscillator frequencies for $N=8$ to realize the effective temperature pulses $\mathcal T(t)$ (plotted in arbitrary units). (b) The temperature of the quantum system depends on the number of harmonic oscillators, $N=4,8,12$. For $N=12$, it comes close to the time-dependent temperature obtained from a Lindblad equation for the quantum system only with $\mathcal{T}(t)$ inserted (black)\revision{, see Eq.~(\ref{eq:effLind})}. The top and bottom curves have been displaced vertically, and we have used $\alpha=0.05\nu\omega_s$, $\gamma=0.01\omega_s$, and $k_B T_e=5\hbar \omega_s$.}
	\label{fig:figure2}
\end{figure}

\revisiontwo{To illustrate our proposal, we now consider the particular case, where the rates $\gamma_j$ are constant, corresponding to a heat bath with a flat spectral density \cite{Chu2017,Hong2017,Wollack2022}. Also, for the sake of simplicity, we take the same rate for all oscillators, $\gamma_j=\gamma$, although none of these assumptions are essential for our proposal.} In the general case of a time-dependent effective temperature, $\mathcal{T}(t)\neq T_e$, we find the time-dependence of the oscillator frequencies by solving Eq.~\eqref{eq:frequencies} for $\omega_j(t)$. For large temperatures, $k_BT_e, k_B\mathcal T (t) \gg\hbar\omega_j(t)$, and small variations, $\mathcal T (t) \equiv T_0 + \Delta (t)$, we find the expression
\begin{equation}
\label{eq:solutionapprox}
    	\frac{\omega_j(t)}{\omega_j(0)}\simeq e^{(1-T_e/T_0) \gamma t }+ \int\limits_0^t \mathrm d \tau \left(\gamma T_e \Delta(\tau) + T_0 \dot \Delta (\tau) \right)/T_0^2.
\end{equation}
Here, as a special case, the integral vanishes for a constant effective temperature, $\mathcal T (t) = T_0$, and we are left only with the first term, which is exponentially decreasing (increasing) if the effective temperature is lower (higher) than the environment temperature. According to Eq.~(\ref{eq:solutionapprox}), the ratio of the frequencies is constant in time, $\omega_j(t)/\omega_i(t)=\omega_j(0)/\omega_i(0)$, and it will be convenient to fix the ratio of adjacent frequencies as $\omega_j(0)/\omega_{j-1}(0)=e^\nu$, where $\nu$ determines their spacing. The initial oscillator frequencies are chosen so that some of them are always close to resonance with the system. The state in Eq.~\eqref{eq:effectivethermal} at the initial time $t=0$ can be produced by first fixing the frequencies to be $\omega_i(0)$ and then letting the harmonic oscillators equilibrate with the environment. Subsequently, the frequencies are changed to $\omega_i(0^+)= \omega_i(0) \mathcal{T}(0^+)/T_e$.

\section{Temperature pulses}
We now apply the effective temperature pulses indicated in brown in Fig.~\ref{fig:system}(b) with $\mathcal T (t) = T_0[1+\lambda\cos(\Omega t)]$ to the left and a step-like profile to the right and then solve Eq.~(\ref{eq:lindblada}) with the time-dependent frequencies from Eq.~(\ref{eq:frequencies}) inserted. To this end, we exploit the fact that Eq.~(\ref{eq:lindblada}) preserves the Gaussian nature of equilibrium states, which are fully described by their correlation matrix $\langle \hat a^\dagger_k \hat a_l\rangle$ with the equation of motion (see App.~\ref{appC})
\begin{equation}
    \frac{\mathrm d }{\mathrm d t} \langle \hat a_k^\dagger \hat a_l\rangle=\sum_{m=0}^N\left( W_{km}\langle \hat a_m^\dagger \hat  a_l\rangle+\langle \hat a_k^\dagger \hat a_m\rangle W^*_{ml}\right)+F_{kl},
    \label{lyaponov}
\end{equation}
which follows from the Lindblad equation for the coupled quantum system and the harmonic oscillators~\cite{adesso2014continuous}. Here, we have defined the $(N+1)\times(N+1)$ block matrix 
\begin{equation}
	\boldsymbol{W}(t)=i \begin{pmatrix}
		\omega_s & \boldsymbol{\alpha}\\
		\boldsymbol{\alpha}^\dagger & \boldsymbol{\Omega}(t)
	\end{pmatrix},
\end{equation}
whose first element contains the frequency of the system oscillator, and the row matrix $\boldsymbol\alpha=\alpha(1,1,\dots, 1)$ describes the coupling between the quantum system and the  harmonic oscillators. We have also defined the diagonal matrix, $\Omega_{kl}(t)=\delta_{kl} (\omega_k(t) +i\gamma/2)$, with the $N$ oscillator frequencies and the damping strength, while $\boldsymbol{F}$ is a diagonal matrix with 0 as the first element followed by $\gamma n_{T_e}(\omega_k)$ for each of the $N$ oscillators. We note that $\langle \hat a_k^\dagger\rangle=\langle \hat a_k\rangle =0$ for all oscillators. The system oscillator is always in a thermal state with the effective temperature given by the average occupation as $k_BT_s = \hbar\omega_s/ \ln\left(1+1/\langle \hat a_s^\dagger \hat a_s \rangle\right)\simeq \hbar\omega_s\langle \hat a_s^\dagger \hat a_s \rangle$ for $\langle \hat a_s^\dagger \hat a_s \rangle\gg1$ in agreement with the equipartition theorem. 

Figure~\ref{fig:system}(b) shows the temperature of the quantum system in blue in response to the effective temperature pulses implemented with $N=16$ harmonic oscillators. These results are based on solving Eq.~(\ref{lyaponov}) numerically. After a short transient, the temperature of the quantum system adapts to the periodic temperature drive and starts oscillating around the average effective temperature $T_0$.  Since the driving is faster than the equilibration time of the system, the temperature oscillations are smaller than those of the effective temperature, and the phase lags behind the drive~\cite{Potanina:2017,portugal2021heat}. By contrast, if the drive was adiabatic, the system oscillator would all the time have the temperature of the harmonic oscillators. 

\begin{figure}
    \centering
    \includegraphics[width=0.95\columnwidth]{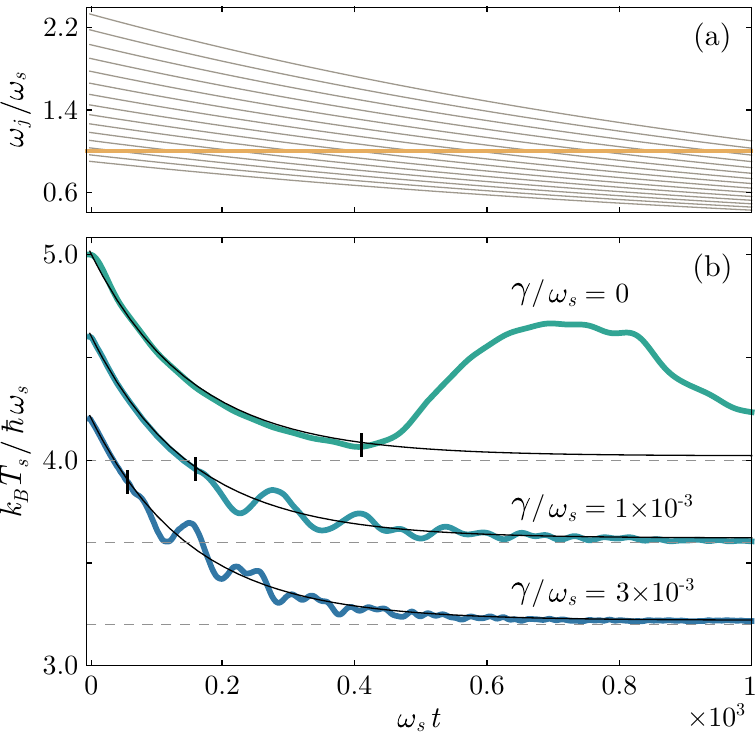}
    \caption{Cooling scheme. (a) The oscillator frequencies are compressed to cool the system.  (b) The system is cooled to the temperature $k_BT_0=4\hbar \omega_s$ below that of the environment, $k_B T_e=5\hbar \omega_s$ using $N=15$ oscillators and  $\alpha=0.05\nu\omega_s$ and  $\gamma=0,0.001\omega_s,0.003\omega_s$ for the three curves with the two lowest being displaced vertically.  We also show the system temperature obtained from a Lindblad equation with the time-dependent temperature $\mathcal T(t)$ inserted (black)\revision{, see Eq.~(\ref{eq:effLind})}. The markers indicate the times $t^*$, where the result are expected to deviate from the Lindblad equation (see text). }
    \label{fig:fig3}
\end{figure}

To corroborate our results, we consider a Lindblad equation for the quantum system only reading
\begin{equation}
	\frac{\mathrm d \hat \rho_s (t)}{\mathrm d t}= -\frac{i}{\hbar}[\hat H_s,\hat \rho_s(t)]+\frac{2\pi }{ \omega_s}\frac{\alpha^2}{\nu^2} \mathcal D_s(t)\hat \rho_s(t),
\label{eq:effLind}
\end{equation}
with the temperature drive $\mathcal T(t)$ entering via the Bose-Einstein distributions in the dissipator $\mathcal D_s(t)$. The prefactor $2\pi (\alpha/\nu)^2/\omega_s$ is related to the spacing of the oscillator frequencies, $\omega_j(0)/\omega_{j-1}(0)=e^\nu$, which ensures a roughly constant density of states around the frequency of the quantum system at all times. By contrast, the prefactor would be time-dependent for other spacings of the frequencies. In Fig.~\ref{fig:system}(b), \revision{we show with black lines the system temperature calculated based on Eq.~(\ref{eq:effLind})} and find good agreement with the solution of Eq.~(\ref{lyaponov}) in blue.  

Figure~\ref{fig:system} demonstrates that effective temperature pulses can be realized using $N=16$ harmonic oscillators. It is, however, interesting to explore what happens if fewer oscillators are used. In Fig.~\ref{fig:figure2}, we show results obtained with $N=4,8,$ and $12$ oscillators together with the time-dependent frequencies for $N=8$, which are chosen around the system frequency to ensure an efficient transfer of energy between the quantum system and the harmonic oscillators. Surprisingly, merely $N=4$ oscillators are sufficient to mimic the effects of a time-dependent temperature, \revision{even if the quantum system might act back on the oscillators, causing them to deviate from the thermal state in Eq.~(\ref{eq:effectivethermal}). By contrast,} with $N=8$ and $N=12$ oscillators, the agreement with the results of Eq.~(\ref{eq:effLind}) is very good. 

\section{Cooling scheme}
Our setup can also be used to cool the quantum system to a constant temperature below that of the environment, $\mathcal T(t)=T_0 < T_e$. To this end, we compress the frequencies as $\omega_j(t)=\omega_j(0) e^{(1-T_e/T_0  ) \gamma t }$ as illustrated in Fig.~\ref{fig:fig3}(a), and in the special case, where the oscillators are decoupled from the environment ($\gamma=0$), we immediately  quench the frequencies to $\omega_j(t)=\omega_j(0) T_0 /T_e$ for $t>0$, where $T_e$ is the initial temperature. The resulting time-dependent temperature of the quantum system is shown in Fig.~\ref{fig:fig3}(b), where we see clear cooling effects as the temperature approaches the effective temperature of the oscillators. However, it is also obvious that the system cannot be kept cold indefinitely. To ensure an efficient removal of energy from the quantum system, it has to be on resonance with the oscillators, but eventually their frequencies are all reduced below the system frequency as seen in Fig.~\ref{fig:fig3}(a). Moreover, without a reservoir, $N$ harmonic oscillators can only mimic an environment up to the time $t^*=2 (N-1)^2/N\Delta \omega$, where $\Delta \omega$ is the difference between the largest and smallest oscillator frequencies~\cite{PhysRevB.92.155126}, which we here evaluate at the initial time $\Delta \omega=\Delta \omega(0^+)$. These times are indicated in Fig.~\ref{fig:fig3}(b), and they coincide with the time at which we see deviations from the temperature based on Eq.~(\ref{eq:effLind}). In particular, without the environment ($\gamma=0$), the dynamics is reversible, and eventually heat flows back into the quantum system, causing it to heat up. This back-flow is avoided by coupling the oscillators to the environment, as we observe for the two other curves with $\gamma>0$, which eventually reach the desired temperature $T_0$.

\begin{figure}
    \centering
    \includegraphics[width=0.95\columnwidth]{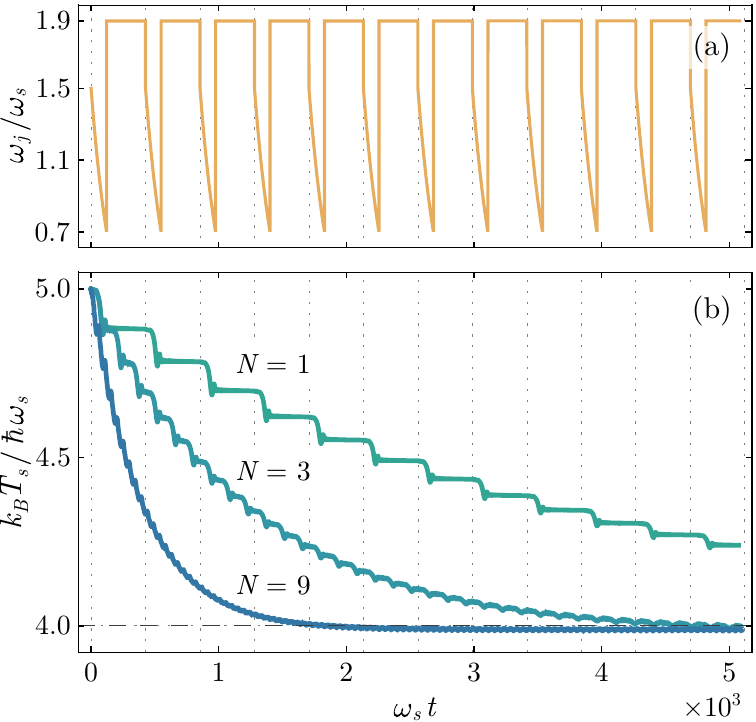}
    \caption{Resetting protocol. (a) We show one of the frequencies in the resetting protocol. (b) Gradual cooling of the quantum system with $N=1, 3, 9$ harmonic oscillators, and $\gamma=0.025\omega_s$, $\alpha= 0.01\omega_s$, $k_BT_0= 4\hbar \omega_s,$ and $k_B T_e=5\hbar \omega_s$.}
    \label{fig:figure3}
\end{figure}

Finally, we extend our cooling scheme by carrying  out the cooling process in several steps as illustrated in Fig.~\ref{fig:figure3}. Figure~\ref{fig:figure3}(a) shows how we use each oscillator to cool the quantum system by first tuning the oscillator across the frequency of the quantum system, so that it cools down, and subsequently we increase the frequency so that it is away from resonance, and the oscillator is then reset by interacting with the environment before the next cooling step. By doing so periodically, we can cool the quantum system in a stepwise manner as shown in Fig.~\ref{fig:figure3}(b) for $N=1,3,9$ oscillators. In particular, for $N=9$, we observe a smooth cooling curve.

\section{Experimental perspectives}
Our scheme can be realized in a variety of experimental setups involving quantum harmonic oscillators, for example, mechanical nano-resonators \cite{Chu2017,Hong2017,Wollack2022,OConnell2010,Chan2011}, microwave cavities~\cite{Wallraff2004,Kurpiers2018}, or electromagnetic resonators~\cite{Hofheinz2008,Hofheinz2009}. With typical frequencies in the range $\omega_s \simeq  1-10$ GHz, the associated temperatures and timescales are about $T\simeq 50-500$ mK and $t\simeq 10^3/\omega_s\simeq 0.1-1$~$\mu$s, which are reachable in current low-temperature experiments. While we here have focused on quantum harmonic oscillators, we note that the scheme can also be implemented with fermionic two-level systems using Eq.~\eqref{eq:frequencies} for the frequency splittings, however, with the Bose-Einstein distributions replaced by Fermi-Dirac distributions. Also, the open quantum system itself does not have to be a harmomic oscillator; rather, it can be any system with discrete energy levels. 

\section{Conclusions and outlook}
We have proposed and analyzed a quantum mechanical scheme to control the effective temperature of an open quantum system. Specifically, by controlling the frequencies of a collection of harmonic oscillators that mediate the interactions between a quantum system and a thermal reservoir, we can implement arbitrary effective temperature drives that the quantum system experiences. We can thereby drive the quantum system with periodic temperature pulses, or it can be cooled to temperatures below that of the environment. Surprisingly, the scheme can be realized with only a few harmonic oscillators, and given the general nature of our proposal, we believe that it can be realized in a variety of experimental setups with the aim to perform coherent manipulations of small quantum systems. Our scheme may also be important for experiments that explore the foundations of quantum thermodynamics. \revision{Finally, our scheme may be extended in many directions, for example, to systems with memory effects~\cite{Breuer2016}.}

\begin{acknowledgements}
\emph{Acknowledgements.}--- We acknowledge support from  Academy of Finland through the Finnish Centre of Excellence in Quantum Technology (Projects No.~312057 and No.~312299) and Grants No.~318937 and No.~331737.
\end{acknowledgements}

%
%
%
%
%
%
%
%
%
\appendix
\onecolumngrid{
\section{Derivation and validity of Equation~(\ref{eq:lindblada})}
\label{appA}	
	\noindent Here, we derive Eq.~(\ref{eq:lindblada}), inspired by the derivation in Ref.~\cite{Rivas:2012} for a constant Hamiltonian. To extend the discussion to a time-dependent Hamiltonian, we follow the derivation in Ref.~\cite{Albash:2012}, which we adapt to our specific setup. In doing so, we identify the particular conditions for Eq.~(\ref{eq:lindblada}) to hold, which we indicate in boxes below.\\
	
	\noindent Our starting point is the total Hamiltonian
	\begin{equation}
		\hat H_T(t)=\hat H(t)+\hat H_B+\hat H_I,
	\end{equation}
	which describes several coupled oscillators. The time-dependent term is given by Eq.~(\ref{eq:Ham}) and reads
	\begin{equation}
		\hat H(t) = \hbar \omega_s\left(\hat a_s^\dagger \hat a_s +1/2\right)+\sum_{j=1}^N \hbar \omega_j(t) \left (\hat a_j^\dagger \hat a_j+1/2\right)+\hbar \alpha \sum_{j=1}^N \left(\hat a_s^\dagger \hat a_j + \hat a_j^\dagger \hat a_s\right),
	\end{equation}
	with $\hat a^{(\dagger)}_{s,j}$ being the usual ladder operators. The first oscillator (with frequency $\omega_{s}$) is the system oscillator, while we refer to the others (with frequencies $\omega_j(t)$) as the ancilla oscillators. The coupling between them is denoted by $\alpha$. We model the thermal bath  with a Hamiltonian that describes a continuous spectrum of harmonic oscillators,
	\begin{equation}
		\hat H_B = \hbar\int_0^{\omega_{m}} d\omega \hat b^\dagger_\omega \hat b_\omega,
	\end{equation}
	where $\hat b^{(\dagger)}_\omega$ are the ladder operators of the bath oscillators, and $\omega_{m}$ is a cut-off frequency.  Finally, the interaction between the ancilla oscillators and the bath is described by the interaction Hamiltonian
	\begin{equation}
		\hat H_I=\hbar g \sum_{j=1}^N \int_0^{\omega_{m}} d\omega h_j[\omega] (\hat a_j^\dagger \hat b_\omega + \hat a_j \hat b^\dagger_\omega),
		\label{Interaction Hamiltonian}
	\end{equation}
	where $g$ is the interaction strength and $h_j[\omega]$ is a  function that describes the coupling to each oscillator.\\
	
	\noindent The full system-ancilla-bath density matrix evolves according to the Liouville-von Neumann equation
	\begin{equation}
		\frac{d}{dt}\hat \rho_T(t)=-\frac{i}{\hbar}[\hat H_T(t),\hat \rho_T(t)].
	\end{equation}
	Our aim is to describe the system-ancilla dynamics itself, taking into account the coupling to the thermal bath.  To this end, we switch to the interaction picture with respect to $\hat H(t)+\hat H_B$ by introducing the unitary operator
	\begin{equation}
		\hat U_0(t,t_0)=\hat{T}\left\{ e^{-i\int_{t_0}^tds[\hat H(s)+\hat H_B]/\hbar}\right\}=\hat{T}\left\{ e^{-i\int_{t_0}^tds \hat H(s)/\hbar}\right\}\otimes e^{-i\hat H_B(t-t_0) /\hbar}\equiv \hat U(t,t_0)\otimes \hat U_B(t,t_0),
	\end{equation}
	where $\hat{T}$ is the chronological time-ordering operator. In the interaction picture, the full density matrix is defined as $\tilde \rho_T(t)= \hat U_0^\dagger(t,0) \hat \rho_T(t) \hat U_0(t,0)$, and its equation of motion easily follows as
	\begin{equation}
		\begin{split}
			\frac{d}{dt}\tilde \rho_T(t)=-\frac{i}{\hbar}[\tilde H_I(t),\tilde \rho_T(t)],
		\end{split}
		\label{eq:eom-rhoT}
	\end{equation}
	with the interaction Hamiltonian in the interaction picture given by $\tilde H_I(t)=\hat U^\dagger_0(t,0) \hat H_I\hat U_0(t,0)$. Equation~(\ref{eq:eom-rhoT}) can be integrated as $\tilde \rho_T(t)=\tilde \rho_T(0)-\frac{i}{\hbar}\int_0^tds [\tilde H_I(s),\tilde \rho_T(s)]$ and iterated once to yield
	\begin{equation}
		\tilde \rho_T(t)=\tilde \rho_T(0)-\frac{i}{\hbar}\int_0^tds [\tilde H_I(s),\tilde \rho_T(0)]-\frac{1}{\hbar^2}\int_0^t ds  \int_0^s ds' [\tilde H_I(s),[\tilde H_I(s'),\tilde \rho_T(s')]].
	\end{equation}
	Differentiating this expression with respect to time, we find
	\begin{equation}
		\frac{d}{dt}\tilde \rho_T(t)=-\frac{i}{\hbar } [\tilde H_I(t),\tilde \rho_T(0)]- \frac{1}{\hbar^2} [\tilde H_I(t),\int_0^t ds [\tilde H_I(t-s),\tilde \rho_T(t-s)]].
		\label{eq:2ndordereom}
	\end{equation}
	Next, we consider the system-ancilla density matrix by tracing out the bath degrees of freedom, yielding
	\begin{equation}
		\tilde \rho(t)=\mathrm{tr}_B\{\hat U_0^\dagger(t,0) \hat \rho_T(t) \hat U_0(t,0)\}=\hat U^\dagger(t,0)\mathrm{tr}_B\{\hat U_B^\dagger(t,0) \hat \rho_T(t) \hat U_B(t,0)\}\hat U(t,0)=\hat U^\dagger(t,0)\hat \rho(t) \hat U(t,0),
	\end{equation}
	where we have used the fact that $\mathrm{tr}_B\{\hat U_B^\dagger(t,0) \hat \rho_T(t) \hat U_B(t,0)\}=\mathrm{tr}_B\{\hat U_B(t,0)\hat U_B^\dagger(t,0) \hat \rho_T(t) \}=\hat \rho(t)$.
	From Eq.~(\ref{eq:2ndordereom}), we then find
	\begin{equation}
		\frac{d}{dt}\tilde \rho(t)=-\frac{i}{\hbar} \mathrm{tr}_B\{[\tilde H_I(t),\tilde \rho_T(0)]\}-\frac{1}{\hbar^2} \mathrm{tr}_B\{[\tilde H_I(t),\int_0^t ds [\tilde H_I(t-s),\tilde \rho_T(t-s)]]\}.
		\label{Equation before Born}
	\end{equation} \\
	Up until this point, we have not made any approximations. However, to proceed, we first apply the standard Born approximation, where correlations between the system and the ancilla, on the one hand, and the bath, on the other hand, are neglected, and we may factorize the density matrix as
	
	\begin{center}{\boxed{1. \qquad \tilde \rho_T(t)=\tilde \rho(t)\otimes \tilde \rho_B(t), \quad \text{with} \quad  \tilde \rho_B(t)= \hat \rho_B = e^{-\hat H_B/(k_BT_e)}/\text{tr}\{e^{-\hat H_B/(k_BT_e)} \}.\qquad \text{(\emph{Born approximation})} \qquad }}\end{center}
	We note that the Born approximation in the context of time-dependent problems is discussed in more detail in Ref.~\cite{Albash:2012}. We have also assumed that the bath remains in its thermal state with a fixed temperature $T_e$. Furthermore, we note that the first term in Eq.~\eqref{Equation before Born} vanishes since
	\begin{equation}
		\mathrm{tr}_B\{[\tilde H_I(t),\tilde \rho_T(0)]\} =\mathrm{tr}_B\{[\tilde H_I(t),\tilde \rho(0)\otimes \hat \rho_B]\} = 0
	\end{equation}
	for the thermal bath state. Together with the explicit expression for the interaction Hamiltonian
	\begin{equation}
		\tilde H_I(t)=\hat U^\dagger_0(t,0) \hat H_I\hat U_0(t,0)=\hbar g \sum_{j=1}^N\int_0^{\omega_{m}} d\omega h_j[\omega] \left[ \tilde a_j^\dagger(t) \tilde b_\omega(t) + \tilde a_j(t) \tilde b^\dagger_\omega(t)\right],
	\end{equation}
	where $\tilde a_j^{(\dagger)}(t) = \hat U^\dagger(t,0) \hat a_j^{(\dagger)} \hat U(t,0)$ and $\tilde b_\omega^{(\dagger)}(t) = \hat U^\dagger_B(t,0) \hat b_\omega^{(\dagger)} \hat U_B(t,0)$ are the ladder operators in the interaction picture, we then obtain from Eq.~\eqref{Equation before Born} the expression
	\begin{equation}
		\begin{split}
			\frac{d}{dt}\tilde \rho(t)=&-g^2 \sum_{j=1}^N\sum_{k=1}^N\int_0^{\omega_{m}} d\omega h_j[\omega]\int_0^{\omega_{m}} d\omega' h_k[\omega']\int_0^t ds  \\
			& \times \mathrm{tr}_B\left\{\left[ \tilde a_j^\dagger(t) \tilde b_\omega(t) + \tilde a_j(t) \tilde b^\dagger_\omega(t),\left[ \tilde a_k^\dagger(t-s) \tilde b_{\omega'}(t-s) + \tilde a_k(t-s) \tilde b^\dagger_{\omega'}(t-s),\tilde \rho(t-s)\otimes \hat \rho_B\right]\right]\right\}.
		\end{split}
		\label{Equation after Born}
	\end{equation}
	Writing out the commutators, we find
	\begin{equation}
		\begin{split}
			\frac{d}{dt}\tilde \rho(t)=g^2 \sum_{j=1}^N\sum_{k=1}^N&\int_0^{\omega_{m}} d\omega h_j[\omega]\int_0^{\omega_{m}} d\omega' h_k[\omega']\int_0^t ds  \\
			\times \bigg( &\Big[\tilde a_k^\dagger(t-s) \tilde \rho(t-s)\tilde a_j^\dagger(t)- \tilde a_j^\dagger(t) \tilde a_k^\dagger(t-s)\tilde \rho(t-s)  \Big]\langle \tilde b_\omega(t)\tilde b_{\omega'}(t-s)\rangle\\
			+& \Big[\tilde a_j^\dagger(t)  \tilde \rho(t-s)\tilde a_k^\dagger(t-s)-\tilde \rho(t-s)\tilde a_k^\dagger(t-s)\tilde a_j^\dagger(t) \Big] \langle \tilde b_{\omega'}(t-s)\tilde b_\omega(t)\rangle \\
			+& \Big[\tilde a_k(t-s) \tilde \rho(t-s)\tilde a_j^\dagger(t)- \tilde a_j^\dagger(t)  \tilde a_k(t-s)\tilde \rho(t-s) \Big] \langle\tilde b_\omega(t)\tilde b^\dagger_{\omega'}(t-s)\rangle  \\
			+&\Big[ \tilde a_j^\dagger(t) \tilde \rho(t-s) \tilde a_k(t-s)- \tilde \rho(t-s) \tilde a_k(t-s) \tilde a_j^\dagger(t) \Big]\langle \tilde b^\dagger_{\omega'}(t-s)\tilde b_\omega(t) \rangle \\
			+&\Big[\tilde a_k^\dagger(t-s)\tilde \rho(t-s)\tilde a_j(t)- \tilde a_j(t) \tilde a_k^\dagger(t-s)\tilde \rho(t-s)  \Big]\langle \tilde b^\dagger_\omega(t) \tilde b_{\omega'}(t-s)\rangle\\
			+&\Big[\tilde a_j(t) \tilde \rho(t-s)\tilde a_k^\dagger(t-s)  -\tilde \rho(t-s)\tilde a_k^\dagger(t-s) \tilde a_j(t) \Big]\langle \tilde b_{\omega'}(t-s) \tilde b^\dagger_\omega(t)\rangle \\
			+&\Big[ \tilde a_k(t-s)\tilde \rho(t-s)\tilde a_j(t)-\tilde a_j(t)  \tilde a_k(t-s)\tilde \rho(t-s) \Big] \langle \tilde b^\dagger_\omega(t) \tilde b^\dagger_{\omega'}(t-s)\rangle\\
			+& \Big[\tilde a_j(t)\tilde \rho(t-s)\tilde a_k(t-s) -\tilde \rho(t-s)\tilde a_k(t-s)  \tilde a_j(t)\Big] \langle \tilde b^\dagger_{\omega'}(t-s) \tilde b^\dagger_\omega(t) \rangle\bigg),
		\end{split}
	\end{equation}
	where all expectation values are evaluated with respect to the equilibrium state of the bath.
	
	Since the state of the bath is assumed to be stationary, the bath correlation functions are translational invariant in time, $\langle \tilde b^{(\dagger)}_\omega(t) \tilde b^{(\dagger)}_{\omega'}(t-s)\rangle=\langle \tilde b^{(\dagger)}_\omega(s) \tilde b^{(\dagger)}_{\omega'}(0)\rangle$. In the following, we let $\tau_B$ denote the characteristic correlation time scale on which the bath correlations $\langle \tilde b^{(\dagger)}_\omega(t) \tilde b^{(\dagger)}_{\omega'}(0)\rangle \sim e^{-t/\tau_B}$ decay. Considering sufficiently long times, $t\gg \tau_B$, and assuming that the system-ancilla relaxation time $\tau_R$ (which in the interaction picture only depends on $g\ll 1$ and not on the internal time scales of the system and the ancilla) is much longer than the relaxation dynamics of the bath, $\tau_R \gg \tau_B$, we can take
	
	\begin{center}{\boxed{2. \qquad \tilde \rho(t-s)\simeq \tilde \rho(t),\qquad \text{(\emph{Markov approximation})} \qquad }}\end{center}
	
	\noindent and extend the upper integration limit to infinity. We then obtain
	\begin{equation}
		\begin{split}
			\frac{d}{dt}\tilde \rho(t)=g^2 \sum_{j=1}^N\sum_{k=1}^N&\int_0^{\omega_{m}} d\omega h_j[\omega]\int_0^{\omega_{m}} d\omega' h_k[\omega']\int_0^\infty ds  \\
			\times \bigg( &\Big[\tilde a_k^\dagger(t-s) \tilde \rho(t)\tilde a_j^\dagger(t)- \tilde a_j^\dagger(t) \tilde a_k^\dagger(t-s)\tilde \rho(t)  \Big]\langle \tilde b_\omega(s)\tilde b_{\omega'}(0)\rangle\\
			+&\Big[\tilde a_j^\dagger(t)  \tilde \rho(t)\tilde a_k^\dagger(t-s)-\tilde \rho(t)\tilde a_k^\dagger(t-s)\tilde a_j^\dagger(t) \Big] \langle \tilde b_{\omega'}(0)\tilde b_\omega(s)\rangle \\
			+&\Big[\tilde a_k(t-s) \tilde \rho(t)\tilde a_j^\dagger(t)- \tilde a_j^\dagger(t)  \tilde a_k(t-s)\tilde \rho(t) \Big] \langle\tilde b_\omega(s)\tilde b^\dagger_{\omega'}(0)\rangle  \\
			+&\Big[ \tilde a_j^\dagger(t) \tilde \rho(t) \tilde a_k(t-s)- \tilde \rho(t) \tilde a_k(t-s) \tilde a_j^\dagger(t) \Big]\langle \tilde b^\dagger_{\omega'}(0)\tilde b_\omega(s) \rangle \\
			+ &\Big[\tilde a_k^\dagger(t-s)\tilde \rho(t)\tilde a_j(t)- \tilde a_j(t) \tilde a_k^\dagger(t-s)\tilde \rho(t)  \Big]\langle \tilde b^\dagger_\omega(s) \tilde b_{\omega'}(0)\rangle\\
			+&\Big[\tilde a_j(t) \tilde \rho(t)\tilde a_k^\dagger(t-s)  -\tilde \rho(t)\tilde a_k^\dagger(t-s) \tilde a_j(t) \Big]\langle \tilde b_{\omega'}(0) \tilde b^\dagger_\omega(s)\rangle \\
			+&\Big[ \tilde a_k(t-s)\tilde \rho(t)\tilde a_j(t)-\tilde a_j(t)  \tilde a_k(t-s)\tilde \rho(t) \Big] \langle \tilde b^\dagger_\omega(s) \tilde b^\dagger_{\omega'}(0)\rangle\\
			+& \Big[\tilde a_j(t)\tilde \rho(t)\tilde a_k(t-s) -\tilde \rho(t)\tilde a_k(t-s)  \tilde a_j(t)\Big] \langle \tilde b^\dagger_{\omega'}(0) \tilde b^\dagger_\omega(s) \rangle\bigg).
		\end{split}
		\label{Intermediate step}
	\end{equation}
	We now express the ladder operators in the interaction picture in terms of the operators in the Schrödinger picture. To this end, we consider the regime of a weak-coupling between the system and the ancilla,
	
	\begin{center}{\boxed{3. \qquad \alpha \ll \omega_s,\omega_j,\qquad \text{(\emph{System-ancilla weak coupling regime})} \qquad }}\end{center}
	
	\noindent such that the ladder operators $\hat a^{(\dagger)}_j$ are (approximate) eigenoperators of the superoperator $\mathcal{H}(t) = [\hat H(t), $ $ \cdot $ $]$, with $\mathcal H(t) \hat a_j^\dagger = \hbar \omega_j(t) \hat a_j^\dagger$ and $\mathcal H(t) \hat a_j = -\hbar \omega_j(t) \hat a_j$. We note that condition 3 is particular to our model, and we work with this condition since the quantum system should only be weakly coupled to its environment. We then obtain
	\begin{equation}
		\begin{split}
			\tilde{a}_j(t) = \hat U^\dagger(t,0)\hat a_j \hat U(t,0)  =&  \hat{T}^*\left\{ e^{i\int_{0}^t ds' \hat H(s')/\hbar}\right\} \hat a_j \hat{T}\left\{ e^{-i\int_{0}^tds' \hat H(s')/\hbar}\right\} =  e^{-i\int_{0}^tds' \omega_j(s')} \hat a_j, \quad \text{and}\\
			\tilde{a}^\dagger_j(t) = \hat U^\dagger(t,0)\hat a_j^\dagger \hat U(t,0)  =&  \hat{T}^*\left\{ e^{i\int_{0}^t ds' \hat H(s')/\hbar}\right\} \hat a_j^\dagger \hat{T}\left\{ e^{-i\int_{0}^tds' \hat H(s')/\hbar}\right\} =  e^{i\int_{0}^tds' \omega_j(s')} \hat a_j^\dagger,
		\end{split}
	\end{equation}
	where $\hat{T}^*$ is the anti-chronological time-ordering operator. Furthermore, for $\tilde{a}_k^{(\dagger)}(t-s)$, we assume that $\hat H(t)$ changes on a time scale $\tau_A$ that is much longer than the time scale $\tau_B$ on which the correlation functions $\langle \tilde b^{(\dagger)}_\omega(t) \tilde b^{(\dagger)}_{\omega'}(0)\rangle$ decay,
	
	\begin{center}{\boxed{4. \qquad \tau_A \gg \tau_B,\qquad \text{(\emph{Slow driving compared with the decay of bath correlations})} \qquad }}\end{center}
	
	\noindent Importantly, this allows us to make the approximation 
	\begin{equation}
		\hat U(t-s,0) =  \hat U^\dagger(t,t-s)\hat U(t,0)\simeq e^{i\hat H(t) s/\hbar}\hat U(t,0),
	\end{equation}
	where we have used the fact that $\hat H(t)$ is approximately constant on the time scale on which the bath correlations decay, $\hat U^\dagger(t,t-s) = \hat{T}^*\left\{ e^{i\int_{t-s}^{t}ds' \hat H(s')/\hbar}\right\}\simeq   e^{i \hat H(t)\int_{t-s}^{t}ds' /\hbar} = e^{i\hat H(t) s/\hbar}$. Using this approximation, we obtain
	\begin{equation}
		\begin{split}
			\tilde{a}_k(t-s) = \hat U^\dagger(t-s,0)\hat a_k \hat U(t-s,0)  \simeq    \hat{T}^*\left\{ e^{i\int_{0}^{t}ds' \hat H(s')/\hbar}\right\} e^{-i\hat H(t)s/\hbar} a_k e^{i\hat H(t)s/\hbar}\hat{T}\left\{ e^{-i\int_{0}^{t}ds' \hat H(s')/\hbar}\right\} \\
			= e^{-i\int_{0}^{t}ds' \omega_k(s')} e^{is \omega_k(t)} a_k, \qquad \text{and}\\
			\tilde{a}_k^\dagger(t-s) = \hat U^\dagger(t-s,0)\hat a_k^\dagger \hat U(t-s,0) \simeq \hat{T}^*\left\{ e^{i\int_{0}^{t}ds' \hat H(s')/\hbar}\right\} e^{-i\hat H(t)s/\hbar} a_k^\dagger e^{i\hat H(t)s/\hbar}\hat{T}\left\{ e^{-i\int_{0}^{t}ds' \hat H(s')/\hbar}\right\} \\
			= e^{i\int_{0}^{t}ds' \omega_k(s')} e^{-is \omega_k(t)} a_k^\dagger. \qquad \hspace{6.2mm}
		\end{split}
	\end{equation}
	Plugging the equations above back into Eq.~\eqref{Intermediate step}, we obtain
	\begin{equation}
		\begin{split}
			\frac{d}{dt}&\tilde \rho(t) =g^2 \sum_{j=1}^N\sum_{k=1}^N\int_0^{\omega_{m}} d\omega h_j[\omega]  \int_0^{\omega_{m}} d\omega' h_k[\omega']\int_0^\infty ds  \\
			\times \bigg[ &\left( \Big[\hat a_k^\dagger \tilde \rho(t)\hat a_j^\dagger- \hat a_j^\dagger \hat a_k^\dagger\tilde \rho(t)  \Big]\langle \tilde b_\omega(s)\tilde b_{\omega'}(0)\rangle +\Big[\hat a_j^\dagger  \tilde \rho(t)\hat a_k^\dagger-\tilde \rho(t)\hat a_k^\dagger\hat a_j^\dagger \Big] \langle \tilde b_{\omega'}(0)\tilde b_\omega(s)\rangle \right)e^{-is\omega_k(t)+i\int_{0}^t ds'[\omega_j(s')+\omega_k(s')]} \\
			+ &\left( \Big[\hat a_k \tilde \rho(t)\hat a_j^\dagger- \hat a_j^\dagger  \hat a_k\tilde \rho(t) \Big] \langle\tilde b_\omega(s)\tilde b^\dagger_{\omega'}(0)\rangle  +\Big[ \hat a_j^\dagger \tilde \rho(t) \hat a_k- \tilde \rho(t) \hat a_k \hat a_j^\dagger \Big]\langle \tilde b^\dagger_{\omega'}(0)\tilde b_\omega(s) \rangle \right)e^{is\omega_k(t)+i\int_{0}^t ds'[\omega_j(s')-\omega_k(s')]}  \\
			+&\left( \Big[\hat a_k^\dagger\tilde \rho(t)\hat a_j- \hat a_j \hat a_k^\dagger\tilde \rho(t)  \Big]\langle \tilde b^\dagger_\omega(s) \tilde b_{\omega'}(0)\rangle +\Big[\hat a_j \tilde \rho(t)\hat a_k^\dagger -\tilde \rho(t)\hat a_k^\dagger \hat a_j \Big]\langle \tilde b_{\omega'}(0) \tilde b^\dagger_\omega(s)\rangle \right) e^{-is\omega_k(t)-i\int_{0}^t ds'[\omega_j(s')-\omega_k(s')]} \\
			+&\left( \Big[ \hat a_k\tilde \rho(t)\hat a_j-\hat a_j  \hat a_k\tilde \rho(t) \Big] \langle \tilde b^\dagger_\omega(s) \tilde b^\dagger_{\omega'}(0)\rangle + \Big[\hat a_j\tilde \rho(t)\hat a_k -\tilde \rho(t)\hat a_k  \hat a_j\right] \langle \tilde b^\dagger_{\omega'}(0) \tilde b^\dagger_\omega(s) \rangle\Big)e^{is\omega_k(t)-i\int_{0}^t ds'[\omega_j(s')+\omega_k(s')]} \bigg].
		\end{split}
	\end{equation}
	We now perform the secular approximation. To this end, we note that for our driving protocols, $\omega_j(t)\neq \omega_k(t)$ for all times, unless $j=k$. Under the assumption that the inverse frequency differences, for all times, are much smaller than the system-ancilla relaxation time, 
	
	\begin{center}{\boxed{5. \qquad |\omega_j(t)-\omega_k(t)|^{-1} \ll \tau_R,\qquad \text{(\emph{Condition for the secular approximation})} \qquad }}\end{center}
	
	\noindent we may neglect all terms except those with a vanishing integral in the exponent, as they oscillate much faster than the typical time scale of the system-ancilla dynamics. This approximation then yields
	\begin{equation}
		\begin{split}
			\frac{d}{dt}\tilde \rho(t)=g^2 \sum_{j=1}^N&\int_0^{\omega_{m}} d\omega h_j[\omega]\int_0^{\omega_{m}} d\omega' h_j[\omega']\int_0^\infty ds  \\
			\times \Bigg(  &\left[\hat a_j \tilde \rho(t)\hat a_j^\dagger- \hat a_j^\dagger  \hat a_j\tilde \rho(t) \right] \langle\tilde b_\omega(s)\tilde b^\dagger_{\omega'}(0)\rangle e^{is\omega_j(t)} +\left[ \hat a_j^\dagger \tilde \rho(t) \hat a_j- \tilde \rho(t) \hat a_j \hat a_j^\dagger \right]\langle \tilde b^\dagger_{\omega'}(0)\tilde b_\omega(s) \rangle e^{is\omega_j(t)} \\
			+&\left[\hat a_j^\dagger\tilde \rho(t)\hat a_j- \hat a_j \hat a_j^\dagger\tilde \rho(t)  \right]\langle \tilde b^\dagger_\omega(s) \tilde b_{\omega'}(0)\rangle e^{-is\omega_j(t)}+\left[\hat a_j \tilde \rho(t)\hat a_j^\dagger -\tilde \rho(t)\hat a_j^\dagger \hat a_j \right]\langle \tilde b_{\omega'}(0) \tilde b^\dagger_\omega(s)\rangle e^{-is\omega_j(t)} \Bigg).
		\end{split}
	\end{equation}
Introducing the bath correlation functions in the frequency domain as
\begin{equation}
	\Gamma_{\omega\omega'}[\omega_j(t)]=\int_0^\infty ds \langle \tilde b_\omega(s) \tilde b_{\omega'}(0)\rangle e^{-is \omega_j(t)} \equiv  \frac{1}{2}\mu_{\omega\omega'}[\omega_j(t)] + i \eta_{\omega\omega'}[\omega_j(t)] ,
	\label{Bath correlation function}
\end{equation}
and the notation $\tilde b_{-\omega}(t)\equiv \tilde b^\dagger_{\omega}(t)$, we obtain
\begin{equation}
	\begin{split}
		\frac{d}{dt}\tilde \rho(t)=g^2 \sum_{j=1}^N & \int_0^{\omega_{m}} d\omega h_j[\omega]\int_0^{\omega_{m}} d\omega' h_j[\omega']  \\
		\times \Bigg( &\left[\hat a_j \tilde \rho(t)\hat a_j^\dagger- \hat a_j^\dagger  \hat a_j\tilde \rho(t) \right]\Gamma_{\omega-\omega'}[-\omega_j(t)] +\left[\hat a_j \tilde \rho(t)\hat a_j^\dagger -\tilde \rho(t)\hat a_j^\dagger \hat a_j \right]\Gamma_{-\omega'\omega}^*[-\omega_j(t)] \\
		+  &\left[\hat a_j^\dagger\tilde \rho(t)\hat a_j- \hat a_j \hat a_j^\dagger\tilde \rho(t)  \right]\Gamma_{-\omega\omega'}[\omega_j(t)] +\left[ \hat a_j^\dagger \tilde \rho(t) \hat a_j- \tilde \rho(t) \hat a_j \hat a_j^\dagger \right]\Gamma^*_{\omega'-\omega}[\omega_j(t)] \Bigg).
	\end{split}
\end{equation}
Separating the Hermitian and non-Hermitian parts of the bath correlation functions, and then transforming back to the Schrödinger picture, we obtain
\begin{equation}
	\begin{split}
		\frac{d}{dt} \hat \rho(t) =& -\frac{i}{\hbar} [\hat H(t) + \hat H_{Ls}(t),\hat\rho(t)] \\
		& +   \sum_{j=1}^N \Bigg[ \gamma^-_j[\omega_j(t)] \left(\hat a_j \hat \rho(t)\hat a_j^\dagger - \frac{1}{2}\left\{ \hat a_j^\dagger \hat a_j,\hat\rho(t) \right\} \right)+\gamma^+_j[\omega_j(t)] \left(\hat a_j^\dagger \hat \rho(t)\hat a_j - \frac{1}{2}\left\{ \hat a_j \hat a_j^\dagger,\hat\rho(t) \right\} \right) \Bigg],
	\end{split}
\end{equation}
where
\begin{equation}
	\gamma^{\pm}_j[\omega_j(t)] = g^2\int_0^{\omega_{m}} d\omega h_j[\omega]\int_0^{\omega_{m}} d\omega' h_j[\omega']  \mu_{\mp \omega\pm\omega'}[\pm \omega_j(t)]
\end{equation}
are the absorption rates ($\gamma^+_j[\omega]$) and the emission rates ($\gamma^-_j[\omega]$) of quanta from/to the thermal bath, and
\begin{equation}
	\hat H_{Ls}(t) = g^2 \sum_{j=1}^N\int_0^{\omega_{m}} d\omega h_j[\omega]\int_0^{\omega_{m}} d\omega' h_j[\omega'] \left( \eta_{\omega- \omega'}[-\omega_j(t)] \hat a^\dagger_{j} \hat a_{j}+  \eta_{-\omega \omega'}[\omega_j(t)] \hat a_{j} \hat a^\dagger_{j} \right)
\end{equation}
is the Lamb-shift Hamiltonian, which commutes with $\hat H(t)$ for $\alpha \ll \omega_s,\omega_j$, and thus only leads to a renormalization of the energy levels. We note that this renormalization is proportional to $g^2\ll1$, and thus we neglect it in the following. In addition, we can derive an explicit expression for the emission rate of quanta to the bath
\begin{equation}
	\begin{split}
		\gamma^-_j[\omega_j(t)] &= g^2\int_0^{\omega_{m}} d\omega h_j[\omega]\int_0^{\omega_{m}} d\omega' h_j[\omega'] \mu_{\omega -\omega'}[-\omega_j(t)] \\
		&= g^2\int_0^{\omega_{m}} d\omega h_j[\omega]\int_0^{\omega_{m}} d\omega' h_j[\omega']\int_{-\infty}^\infty ds e^{is \omega_j(t)} \langle \tilde b_\omega(s) \tilde b^\dagger_{\omega'}(0)\rangle,
	\end{split}
\end{equation}
in terms of the spectral density. Using the fact that $\tilde b_\omega(t) = \hat U^\dagger_B(t,0) \hat b_\omega \hat U_B(t,0)  =  e^{-i\omega s}\hat b_\omega $, we obtain
\begin{equation}
	\begin{split}
		\gamma^-_j[\omega_j(t)] = &g^2\int_0^{\omega_{m}} d\omega h_j[\omega]\int_0^{\omega_{m}} d\omega' h_j[\omega']\int_{-\infty}^\infty ds e^{is [\omega_j(t)-\omega]}\langle \hat b_\omega \hat b^\dagger_{\omega'}\rangle =2\pi g^2\int_0^{\omega_{m}} d\omega' h_j[\omega']h_j[\omega_j(t)]\langle \hat b_{\omega_j(t)} \hat b^\dagger_{\omega'}\rangle\\
		=&2\pi g^2 h_j^2[\omega_j(t)]\langle \hat b_{\omega_j(t)} \hat b^\dagger_{\omega_j(t)}\rangle \equiv2\pi g^2 h_j^2[\omega_j(t)] (1+n_{T_e}[\omega_j(t)]),
	\end{split}
\end{equation}
provided that $0<\omega_j(t) <\omega_{m}$. Here, $n_{T_e}[\omega]$ is the Bose-Einstein distribution, and we have used the fact that $\int_{-\infty}^\infty ds e^{is\omega} = 2\pi \delta(\omega)$, with $\delta(\omega)$ being the Dirac delta function. Introducing the spectral density $J_j[\omega] \equiv 2\pi h_j^2[\omega]$, we find
\begin{equation}
	\begin{split}
		\gamma^-_j[\omega_j(t)] = g^2J_j[\omega_j(t)] (1+n_{T_e}[\omega_j(t)]).
	\end{split}
\end{equation}
Analogously, we find a similar expression for the absorption rate of quanta from the bath reading
\begin{equation}
	\begin{split}
		\gamma^+_j[\omega_j(t)] =  g^2J_j[\omega_j(t)] n_{T_e}[\omega_j(t)].\hspace{8.8mm}
	\end{split}
\end{equation}
We note that the rates fulfill a detailed balance, $\gamma^+_j[\omega_j(t)]/\gamma^-_j[\omega_j(t)] = n_{T_e}[\omega_j(t)]/(1+n_{T_e}[\omega_j(t)]) = e^{-\hbar \omega/(k_B T_e)}$ as expected. Using the expressions for the rates, we finally arrive at the master equation in Eq.~(2) of the main text,
\begin{equation}
	\frac{\mathrm{d} \hat \rho(t)}{\mathrm{d}t} = -\frac{i}{\hbar} [\hat H(t),\hat \rho(t)] + \sum_{j=1}^N \gamma_j[\omega_j(t)]\mathcal{D}_j(t) \hat \rho(t),
\end{equation}
with $\gamma_j[\omega_j(t)]=g^2J_j[\omega_j(t)]$ and the dissipator 
\begin{equation}
	\mathcal{D}_j\hat \rho = \left( 1+n_{T_e}[\omega_j(t)]\right)\left(\hat a_j \hat \rho \hat a_j^\dagger - \frac{1}{2} \{\hat a_j^\dagger \hat a,\hat \rho\}\right)+n_{T_e}[\omega_j(t)] \left(\hat a_j^\dagger\hat \rho \hat a_j - \frac{1}{2}\{ \hat a_j \hat a_j^\dagger,\hat \rho\}\right).
\end{equation}
}
\twocolumngrid

\section{Derivation of Equation~(\ref{eq:frequencies})}
\label{appB}

As our starting point, we take Eq.~(\ref{eq:lindblada}) with $\alpha=0$, for which it for the oscillators simplifies to
\begin{equation}\label{eq:lindblad1}
	\frac{\mathrm d \hat \rho (t)}{\mathrm d t}= -i\sum_{j=1}^{N}  \omega_{j} [\hat a_j^\dagger \hat a_j,\hat \rho(t)] + \sum_{j=1}^{N}\gamma_j(\omega_j)  \mathcal D_j\hat \rho(t),
\end{equation}
having omitted the explicit time-dependence of the oscillator frequencies, $\omega_j=\omega_j(t)$.
In this case, the harmonic oscillators are independent, and we can treat each of them separately as
\begin{equation}\label{eq:lindblada2}
	\frac{\mathrm d \hat \rho_j(t)}{\mathrm d t}= -i \omega_j [\hat a_j^\dagger \hat a_j,\hat \rho_j(t)] +  \gamma_j(\omega_j) \mathcal D_j\hat \rho_j(t), j=1,\ldots,N.
\end{equation}
Next, we introduce a characteristic function for each oscillator by defining
\begin{equation}
	\chi_j(\lambda,t)= \mathrm{Tr} \left(  \hat\rho_j(t)  \hat D_j(\lambda) \right), \quad \hat D_j(\lambda)= e^{\lambda  \hat a_j^\dagger - \lambda^*  \hat a_j },
\end{equation}
where $\hat D_j(\lambda)$ is the displacement operator. The equation of motion for $\chi_j(\lambda,t)$ follows from Eq.~(\ref{eq:lindblada2}) and reads
\begin{equation}
	\label{eq_eom}
	\frac{\mathrm d \chi_j (\lambda,t)}{\mathrm d t}= \mathcal U_j \chi_j(\lambda,t)+\mathcal D_j \chi_j(\lambda,t),
\end{equation}
where 
\begin{equation}
	\mathcal U_j \chi_j(\lambda,t)= -i \omega_j (t)[\lambda^* \partial_{\lambda^*}- \lambda \partial_{\lambda} ]\chi_j(\lambda,t)
\end{equation}
corresponds to the unitary dynamics, while the dissipator leads to the term
\begin{equation}
	\mathcal D_j \chi_j(\lambda,t)= -(\gamma(\omega_j)/2)[\lambda^* \partial_{\lambda^*}+ \lambda \partial_{\lambda}+\lambda^*\lambda(2 n_B+1) ]\chi_j(\lambda,t).
\end{equation}
Now, the characteristic function of a thermal state with average occupation $n_{\mathcal{T}}(t)$ reads
\begin{equation}
\chi_{\mathcal{T}}(\lambda,t)= e^{- \lambda\lambda^* [n_{\mathcal{T}}(t)+1/2]}.
\end{equation}
Inserting this expression into Eq.~(\ref{eq_eom}), we then find 
\begin{equation}
	\begin{split}
	\frac{\mathrm d \chi_{\mathcal{T}} (\lambda,t)}{\mathrm d t}=&-\lambda \lambda^* \dot n_{\mathcal{T}} (t)\chi_{\mathcal{T}} (\lambda,t)\\
	=&\lambda \lambda^*\gamma(\omega_j(t))\left[n_{\mathcal{T}}(t)-n_B \right]\chi_{\mathcal{T}} (\lambda,t),
	\end{split}
\end{equation}
which directly leads us to Eq.~(\ref{eq:frequencies}).

\section{Derivation of Equation~(\ref{lyaponov})}
\label{appC}

We start by writing the Hamiltonian in Eq.~(\ref{eq:Ham}) as
\begin{equation}
	\begin{split}
	\hat H(t)= & \hbar \vec a^\dagger \cdot \begin{pmatrix}
		\omega_s & \boldsymbol{\alpha}\\
		\boldsymbol{\alpha}^\dagger & \boldsymbol{\Omega}(t) \end{pmatrix}\cdot \vec a\\
	&= \hbar \vec a^\dagger \cdot \boldsymbol{H}(t)\cdot \vec a\\
	&=\hbar\sum_{m,n}H_{mn}(t) \hat a_m^\dagger \hat a_n
	\end{split}
\end{equation}
where $\vec a = (\hat a_s, \hat a_1 ..., \hat a_N)^\mathrm{T}$ contains the annihilation operators, and we have defined the matrix~$\boldsymbol{H}(t)$ reading
\begin{equation}
	\boldsymbol{H}(t)= \begin{pmatrix}
		\omega_s & \alpha &  \cdots &  \alpha \\
		\alpha & \omega_1 &    \cdots & 0 \\
		\alpha & 0 &  \ddots & 0\\
		\alpha & 0  & \cdots & \omega_N \\
	\end{pmatrix}.
\end{equation}
The  equations of motion for the correlators, $\langle \hat a_k^\dagger \hat a_l\rangle=\mathrm{tr}\{\hat a_k^\dagger \hat a_l\hat{\rho}\}$, follow from Eq.~(\ref{eq:lindblada})  and read
\begin{equation}
	\begin{split}
		\frac{\mathrm d }{\mathrm d t} \langle \hat a_k^\dagger \hat a_l\rangle=& -\frac{i}{\hbar}\mathrm{tr}\{\hat a_k^\dagger  \hat a_l [\hat H,\hat \rho]\} + \gamma\sum_{j=1}^{N} \mathrm{tr}\{\hat a_k^\dagger  \hat a_l \mathcal D_j\hat \rho\}\\
		=& -i\sum_{m,n}H_{mn}\left(\mathrm{tr}\{\hat a_k^\dagger  \hat a_l \hat a_m^\dagger \hat a_n\hat \rho\}-\mathrm{tr}\{\hat a_m^\dagger \hat a_n\hat a_k^\dagger  \hat a_l \hat \rho\}\right) \\
		&+ \gamma\sum_{j=1}^{N} [1+n_{B}(\omega_j)]\mathrm{tr}\{\hat a_k^\dagger  \hat a_l (\hat a_j \hat \rho \hat a_j^\dagger-\frac{1}{2}\{ \hat a_j^\dagger \hat a_j,\hat \rho\})\}\\
		&+ \gamma\sum_{j=1}^{N} n_{B}(\omega_j)\mathrm{tr}\{\hat a_k^\dagger  \hat a_l(\hat a_j^\dagger \hat \rho \hat a_j-\frac{1}{2}\{\hat a_j \hat a_j^\dagger,\hat \rho\})\}\\
		=& \sum_{m} [(iH_{k m})\langle \hat a_m^\dagger \hat a_l \rangle+\langle \hat a_k^\dagger \hat a_m  \rangle(-i H_{l m})]\\
		&+\gamma [n_B(\omega_l) \delta_{kl}- \langle \hat a_k^\dagger \hat a_l\rangle  ]  \\
		=& \sum_{m}\left( W_{km}\langle \hat a_m^\dagger \hat  a_l\rangle+\langle \hat a_k^\dagger \hat a_m\rangle W^*_{ml}\right)+F_{kl},
	\end{split}
\end{equation}
which is Eq.~(\ref{lyaponov}). Here, we have introduced the matrix $\boldsymbol{\Gamma}=\mathrm{diag}(0,\gamma/2,\gamma/2,\ldots,\gamma/2)$ and defined
\begin{equation}
	\boldsymbol{W}=i \boldsymbol{H}-\boldsymbol{\Gamma}
\end{equation}
together with $\boldsymbol{F}=\mathrm{diag}(0,\gamma n_B(\omega_1),\ldots, \gamma n_B(\omega_N))$.

%
%
%
%
%
%
%

%

\end{document}